# Hollow screw-like drill in plasma using an intense Laguerre–Gaussian laser

Wenpeng Wang , Baifei Shen*, Xiaomei Zhang, Lingang Zhang, Yin Shi, and Zhizhan Xu*

State Key Laboratory of High Field Laser Physics, Shanghai Institute of Optics and Fine Mechanics, Chinese Academy of Sciences, P.O. Box 800-211, Shanghai 201800, China;  These authors contributed equally to this work. *e-mail: bfshen@mail.shcnc.ac.cn; zzxu@mail.shcnc.ac.cn

**With the development of ultra-intense laser technology, MeV ions can be obtained from laser–foil interactions in the laboratory. These energetic ion beams can be applied in fast ignition for inertial confinement fusion, medical therapy, and proton imaging. However, these ions are mainly accelerated in the laser propagation direction. Ion acceleration in an azimuthal orientation has been scarcely studied. In this research, a doughnut Laguerre–Gaussian (LG) laser is used for the first time to examine laser–plasma interaction in the relativistic intensity regime in three-dimensional particle-in-cell simulations. Studies have shown that a novel rotation of the plasma is produced from the hollow screw-like drill of an       mode laser. The angular momentum of particles in the longitudinal direction produced by the LG laser is enhanced compared with that produced by the usual laser pulses, such as linearly and circularly polarized Gaussian pulses. Moreover, the particles (including electrons and ions) can be trapped and uniformly compressed in the dark central minimum of the doughnut LG pulse. The hollow-structured LG laser has potential applications in the generation of x-rays with orbital angular momentum, plasma accelerators, fast ignition for inertial confinement fusion, and pulsars in the astrophysical environment.**

Nowadays, laser intensity can increase up to $10^{22}$ W/cm$^2$[1,2]. Energetic protons have been obtained through different mechanisms, such as target normal sheath acceleration[3-17], radiation pressure acceleration[18-37], collisionless shock acceleration[38-42], breakout afterburner[43,44], and a combination of different mechanisms[43-46]. However, these ions are mainly accelerated in the laser propagation direction. Ion acceleration in azimuthal orientation is scarcely mentioned. A circularly polarized (CP) light may carry the angular momentum[47]. The main reason for this phenomenon is that a CP light carries an orbital angular momentum $\iota_z$ ($\iota_z=\pm 1$) per photon. More than 70 years ago, the mechanical torque created by the transfer of angular momentum of a CP light was first observed in Beth's[47] experiments. However, the small quantities of the optical angular momentum are difficult to detect in the CP light experiments.

A laser with a Gaussian mode, such as Laguerre–Gaussian (LG) mode, also possesses an orbital angular momentum[48]. A linearly polarized (LP) LG laser with a helical wave-front structure has a central phase singularity[49]. The angular momentum produced by such structure is sometimes referred to an orbital angular momentum, which is different from the spin angular momentum produced by the CP laser pulse[47]. LG laser pulse is circularly symmetric in the cross-section with respect to the optical axis [the direction of light propagation, Figs. 1(b) to 1(e)]. The mode of the

LG laser pulse ($LG$) is described by integer indices $l$ and $p$, where $l$ denotes the number of 2 phase cycles around the circumference and $(p + 1)$ denotes the number of radial nodes in the mode profile. This study discusses the $LG$ mode, where $l$ 0 indicates the presence of an azimuthal phase term $\exp(-il\phi)$ in the laser mode. $LG$ laser carries an orbital angular momentum $l$ per photon.

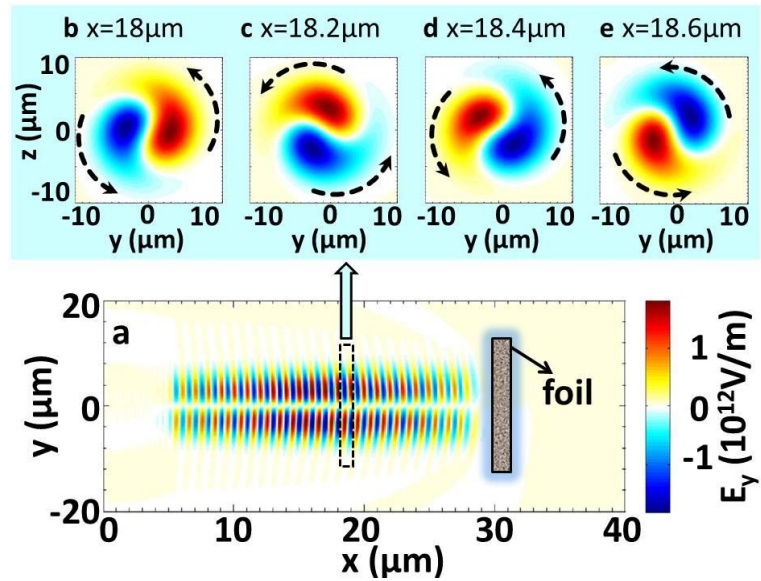

Figure 1  (a) Schematic arrangement of the 3D PIC simulation. The distributions of electric fields in the $(y, z)$ plane at (b) $x = 18$ μm, (c) $x = 18.2$ μm, (d) $x = 18.4$ μm, and (e) $x = 18.6$ μm.

The nature of the orbital angular momentum of different LG modes has been investigated in optical trapping experiments. Allen et al.[48] showed that an LG mode has a well-defined orbital angular momentum. They also observed the torque on suspended cylindrical lenses arising from the reversal helicity of an LG mode. He et al.[50] demonstrated that the absorptive particles trapped in the dark central minimum of a doughnut laser pulse are set into rotation. Furthermore, the rotation particles are controlled using both the spin and the orbital angular momentum of light. In such case, the LG light is beneficial because it reduces the ac stark shift and the broadening of transitions at the trap center. Kuga et al.[51] proposed to trap atoms along the beam center using an LG light. They exploited the spatial profile of LG modes with $p = 0$, which has the form of a ring of light. This feature is important in laser cooling and trapping experiments because the repulsive optical dipole force for blue detuned laser light restricts the atoms to the inner dark region of the laser beam, where photon scattering and the associated heating are minimized. Such hollow-structured LG laser can be used to investigate some difficult problems, such as generation of x-rays with orbital angular momentum[52-54], plasma accelerators[55,56], fast ignition for inertial confinement fusion[57-59], and pulsars in the astrophysical environment[60].

**Results**

In this letter, the doughnut LG laser is used for the first time in relativistic intensity laser plasma interaction. The LG laser rotates electrons and protons, in the azimuthal orientation. Unlike conventional laser pulses, such as LP and CP, enhancement of the proton angular momentum along the longitudinal direction is obtained when an intense LP LG laser pulse irradiates on a thin foil. Three-dimensional (3D) PIC simulations are performed to investigate the LG laser interaction on a foil. In the simulation, the field amplitude $E(\text{LG})$ of an LG laser with mode $(l, p)$ is given by

$$E\left(\text{LG}_p^l\right) = E_0 (-1)^p \exp\left[-\frac{r^2}{W_\perp^2} - \frac{(W_x - ct)^2}{W_x^2}\right]\left(\frac{r\sqrt{2}}{W_\perp}\right)^l L_p^l\left(\frac{2r^2}{W_\perp^2}\right)$$
$$\cos\left[kx - \omega t + \frac{kr^2 x}{2(x_r^2 + x^2)} + (2p + l + 1)\arctan\left(\frac{x}{x_r}\right) + l\phi\right], \quad (1)$$

where $E_0$ is the peak amplitude of the electric field, $r$ is the radius, $W$ is the radius at which the Gaussian term falls to 1/e of its on-axis value, $W_x$ is the pulse length in the $x$ direction, $0 < t < 2W_x/c$, is the generalized Laguerre polynomial, $k$ is the wave number, is the laser frequency, $x$ is the distance from the beam waist, $x_r$ is the Rayleigh range, $x/x_r$ is the Guoy phase of the mode, and $\phi$ is the azimuthal angle[61]. This study mainly discusses the mode of LG, and thus, $p = 0$ and $l = 1$ are used in Eq. (1).

Fig. 2 shows the total angular momentum of the particles (electrons and protons) in 3D PIC simulations. The detailed simulation parameters are shown in the **Methods** section. To describe the rotation effects of LG mode on the plasma, the angular momentum of the particles in the $x$ direction (the longitudinal direction) $m_e(yp_z - zp_y) + m_p(yp_z - zp_y)$ is calculated, where $p_{y,z} = v_{y,z}$ is the velocity and $= (1 - )^{-1/2}$ is the relativistic factor. Fig. 2 further shows the simulation results at different times. The LG laser irradiates on the foil at $t = 30T$, and is totally reflected by the foil at $t \sim 54T$. The angular momentum of the electrons and protons increases up to $-1.55 \times 10^{-17}$ kg·m²/s until the laser pulse is totally reflected. Thus, the trapped particles rotate by the helicity of the LG laser.

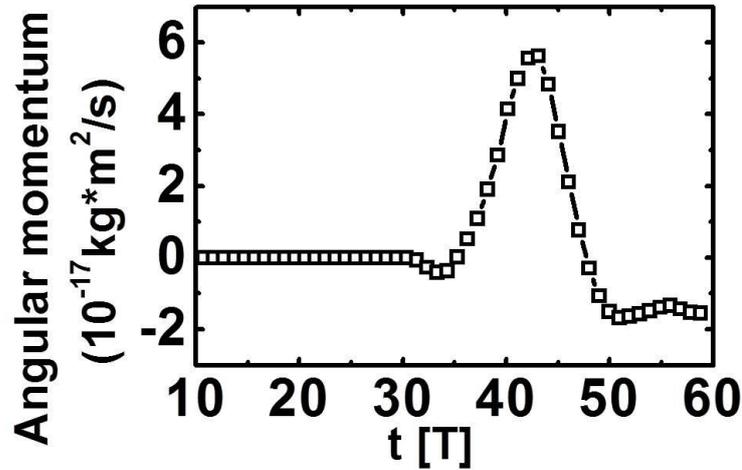

Figure 2     Total angular momentums me(ypz - zpy)+ mp(ypz - zpy) in the x direction.

Fig. 3 shows the angular momentums of particles at $60T$ for different laser amplitudes $a_0$. The radius $W$ and duration $\tau_L$ of the laser pulse remain constant for different $a_0$. The angular momentums increase to $-1.32 \times 10^{-16}$ kg·m$^2$/s for $a_0 = 14$. From Fig. 3, a critical condition for the rotation of the particles is observed at $a_0$ 5. Detailed simulations have proved this condition. The LG laser is almost reflected for $a_0 = 1$ at $t = 50T$ (Fig. 4). The laser transmits through the foil at $a_0 = 5$, and more portions of the laser pulse transmit at $a_0 = 10$. Thus, the critical condition for the proton rotation may be related to the transmission of the foil,[18,35,36,62-65] and may be expressed by $a_0$ $nd$,[35-37] where the foil density $n$ is normalized by $n_c$ and the foil thickness $d$ is normalized by laser wavelength . The theory estimates that the laser starts to transmit through the foil at $a_0 \sim 6.28$ for $n = 2$ and $d = 1$ according to $a_0 \sim nd$, which is larger than the simulation result in Figs. 4(d) to 4(f), where the laser begins to transmit through the foil at $a_0 = 5$. This phenomenon is attributed to the enhancement of the transmission of the laser pulse caused by the self-focusing of the laser pulse and multi-dimensional instabilities, as demonstrated in our previous simulation study[35,36]. Fig. 3 shows that the angular momentums of particles increase when the LG laser transmits through the foil. The main reason is that more laser energy is absorbed by the foil when the laser transmits through the foil. In realistic interaction, the laser transmission may be realized by LG laser pulses with larger intensity and a larger angular momentum of the foil can be generated, similar to the case in Fig. 3.

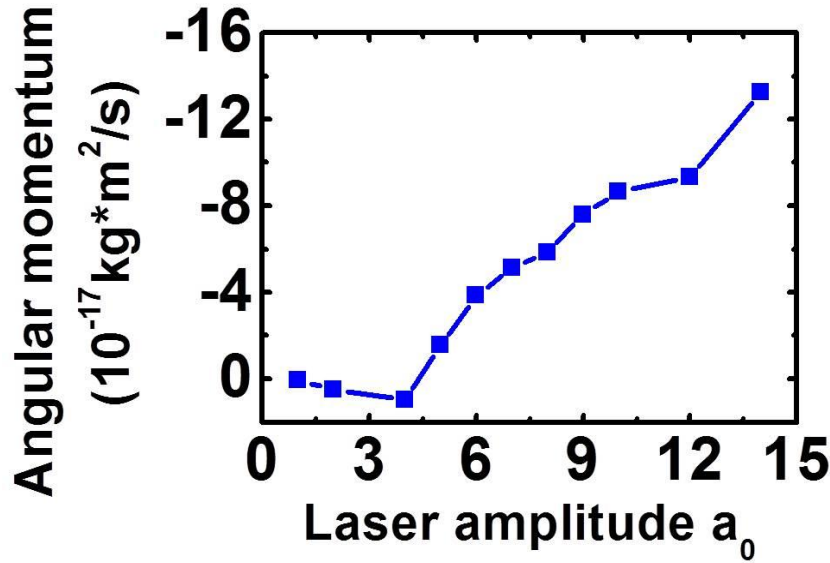

Figure 3    Angular momentums of particles at $60T$ for different laser amplitudes $a$. The initial foil density is $n_0 = 2\ n_c$ and the foil thickness is $d = 1$ μm.

Fig. 5 shows the distribution of electric fields, electron densities, and ion densities in the $(y, z)$ plane inside the foil. The corresponding distributions in the $(x, y)$ plane are shown in Figs. 4(g) to 4(i). The particles (electrons and protons) are rotated in the direction of the helicity of the beam at different position in the foil, just as shown in Fig. 5. The electrons in a ring are drilled out while a compressed point remains at the center at $t = 50T$ [Figs. 5(d) to 5(f)]. Such ring structure is related to the helical structure of the LG mode laser [Figs. 5(a) to 5(c)]. The trapping and compression of the foil at $x = 30.5$ μm are presented at different time window in Fig. 6, which clearly show the process of trapping and compression with time. Fig. 6(a) shows that the electrons are first dragged along the tangential direction at $t = 35T$. The protons remains almost at rest due to their large mass [Fig. 6(e)]. The protons begin to be accelerated by the charge separation electric field between the electrons and protons as LG laser continues to rotate in the foil [Figs. 6(f)-6(h)]. At $t = 50T$, both the electrons and protons are trapped and compressed into one point by the hollow-structured LG laser. In addition, some ripples are generated at the edge of the ring structure [Figs. 6(d) and 6(h)], which confirms that the LG mode laser drills in the plasma like a screw. Such hollow screw-like drill can uniformly trap and compress the plasma at the center [Figs. 5(e) and 5(h)], which may realize the screw-like drilling in the inertial confinement fusion and laser-driven particle accelerations.

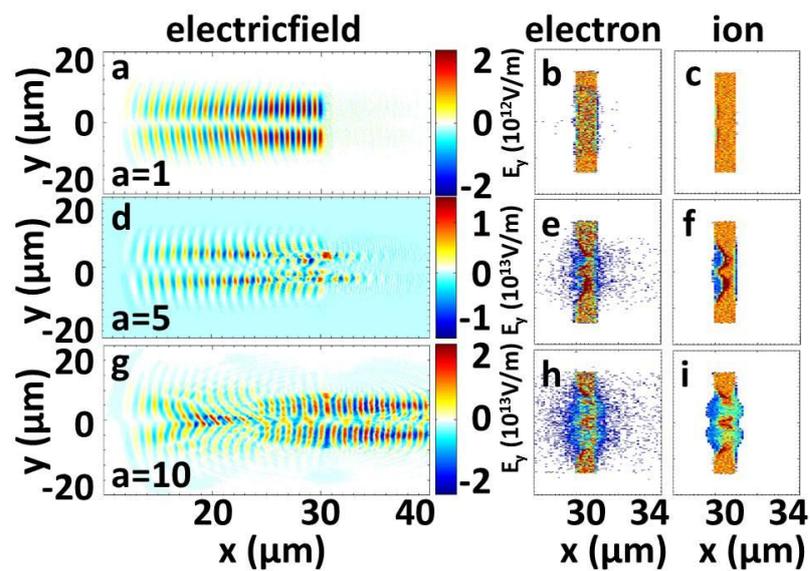

Figure 4  LG laser interactions on the foils for (a–c) $a = 1$, (d–f) $a = 5$, and (g–i) $a = 10$. The distributions of the electric fields (first column), electron density (middle column), and ion density (third column) are shown at $t = 50T$.

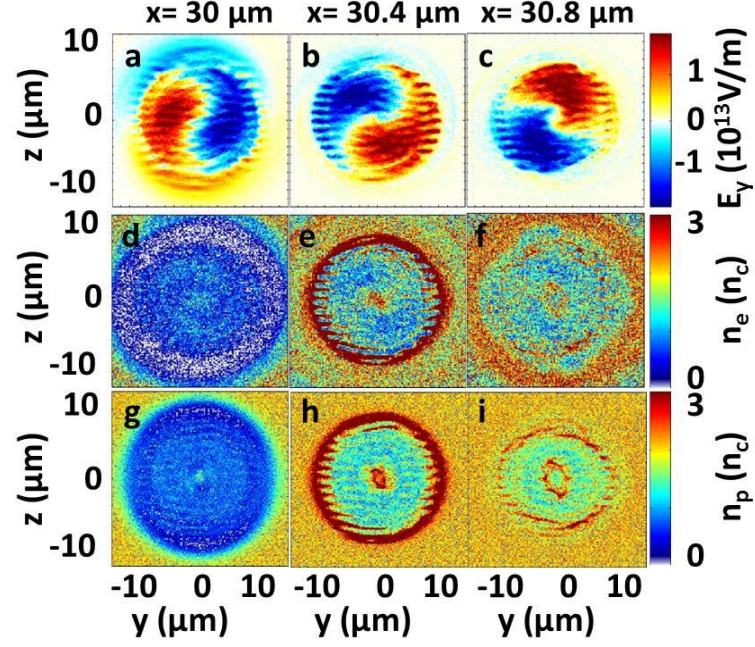

Figure 5    Distributions of (a–c) electric fields, (d–f) electron densities, and (g–i) ion densities in the (y, z) plane at $x = 30$ μm (first column), $x = 30.4$ μm (second column), and $x = 30.8$ μm (third column) at $t = 50T$. The corresponding distributions are shown in Figs. 4(g) to 4(i).

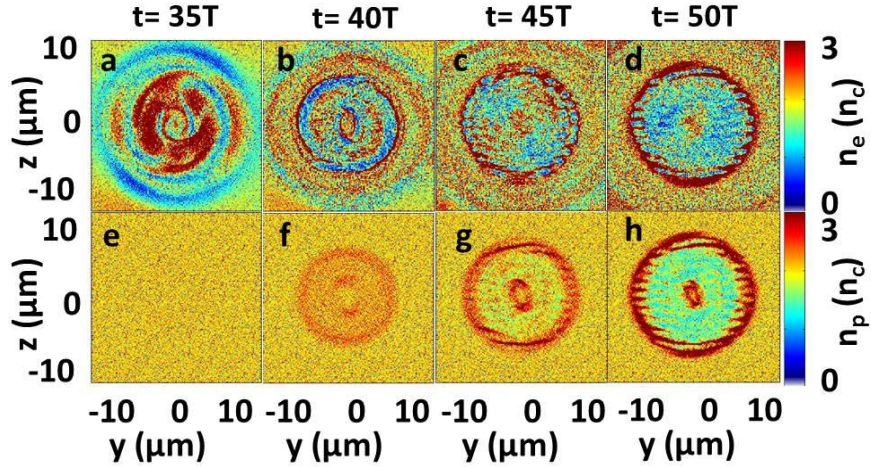

Figure 6    Distributions of (a–d) electron densities and (e–h) ion densities in the (y, z) plane at $x = 30.5$ μm for different time $t = 35T$, $40T$, $45T$, and $50T$.

**Discussion**

The particles are rotated by the LG laser (Fig. 2). The transfer of angular momentum from the laser to the particles is then theoretically estimated in realistic cases. In terms of quantum mechanics, the rotation can be caused by the angular momentum of photons. The LG mode can be seen as the eigenmode of the angular momentum operator of $L_z$[48] and carries an orbital angular momentum of $l$ per photon. The angular momentum carried by a photon of a polarized LG mode laser is $(l+\sigma_z)\hbar$, where $\sigma_z$ is ±1 for the CP laser and 0 for the LP laser. The total angular momentum absorbed from the laser can then be approximately expressed as

$$P_{angular} = \eta \frac{\left(a_0^2 \times 1.37 \times 10^{18} \text{W/cm}^2\right) S \tau_L}{h\nu} (l+\sigma_z)\hbar, \quad (2)$$

where $\eta$ is the absorbing ratio from the laser pulse during the interaction, $S \sim \pi W^2$ and $W$ are the radius at which the Gaussian term falls to $1/e$ of its on-axis value, $h=6.63\times10^{-34}$ J·s is Plank constant, $\nu = 1/T$ is the frequency of light, $h\nu$ is the energy of single photon, and $\tau_L$ is the pulse duration. $P_{angular}=8.22\times10^{-16}$ kg·m²/s is then obtained for an LP LG laser ($l=1$ and $\sigma_z=0$) with $a_0=10$, $W=4$ μm, and $\tau_L=40$ fs. A ring of the target (the radius of the inner and outer ring is 2 and 4 μm, respectively) is assumed to be rotated by the hollow structure of the LG mode laser. Assuming that the angular momentum of the laser is totally transferred to the foil ($\eta=1$), $P_{angular}= \gamma m \omega r^2 \pi d \rho =8.22\times10^{-16}$ kg·m²/s is obtained, where the foil density is $2\times10^3$ kg/m³, foil thickness is $d=1$ μm, $\gamma$ is the relativistic factor, $\omega$ is the angular velocity of the foil, and $r$ is the radius of the foil. Afterward, $\omega \approx 1.09 \times 10^{12}$ rad/s is obtained and the velocity of the particle at $r = 4$ μm is approximately $4.36\times10^6$ m/s. The angular momentum is proportional to the laser amplitude $a_0$ and laser duration $\tau_L$, indicating that the angular momentum absorbed from the LG laser can be increased to a certain extent using a high-intensity long pulse based on Eq. (2). Notably, the increase of the angular momentums of particles mainly depends on the laser energy and the absorbing ratio from the laser pulse during the interaction $\eta$, as shown in Eq. (2). Clearly, a more accurate measurement is taken into account, which states that the total angular momentum is divided by the laser energy. $P_{angular}/P_{laser} \propto \eta$ is obtained when the laser energy is considered as $P_{laser}= S\tau_L$. Previous studies have shown that hole boring is deeper for a higher $a_0$, and the time of hole boring is longer for a larger $\tau_L$ [36,37]. The particles can absorb more energy from the laser pulse (corresponding to a larger $\eta$) with the enhancement of the hole boring. Then, a larger $P_{angular}$ is obtained according to Eq. (2). It should be noted that the total angular momentum transferred from laser to particles can be calculated from Eq. (2) when the laser pulse does not transmit the foil. In this case, the angular momentum $P_{angular}$ is proportional to the laser energy absorbed by the foil ($\eta S\tau_L$). Less angular momentum may be generated when the foil is totally destroyed. In this case, a smaller $\eta$ is obtained and less energy of the LG laser is contributed to total angular momentum of protons.

To show the difference of LG laser on the rotation of the particles, LP and CP laser pulses are also considered. The total angular momentum density per unit power has been defined by Allen *et al.*, where the cases of LP ($\sigma_z=0$) and CP ($\sigma_z=\pm1$) are considered[48]. Fig. 7 shows that the angular momentums of the particles in the longitudinal direction $m_e(yp_z - zp_y)+ m_p(yp_z - zp_y)$ for LG, LP,

and CP laser pulses. The amplitude of the LP pulse is
$a = a_0 \exp\left[-r^2/W_\perp^2 - (W_x - ct)^2/W_x^2\right]\sin(\omega_L t)\hat{\mathbf{e}}_y$ ($a_0$=7.4), and the amplitude of the CP pulse is $a = a_0 \exp\left[-r^2/W_\perp^2 - (W_x - ct)^2/W_x^2\right]\left[\sin(\omega_L t)\hat{\mathbf{e}}_y + \cos(\omega_L t)\hat{\mathbf{e}}_z\right]$ ($a_0$=5.2), where $0 < t < 2W_x/c$. The amplitude of the LG pulse is expressed by Eq. (1), where $a_0$=10, $W$ = 4 μm, and $W_x$ = 12 μm; $d$ = 1 μm, and $n_0$ = 4 $n_c$ are used in three cases. The values of laser energy in three cases are similar. Compared with LP and CP lasers, the LG laser can generate larger angular momentum of particles (Fig. 7), indicating that the LG laser is beneficial for the rotation of the particles. Compared with case of CP, the doughnut-structured LG pulse (see Fig. 1) has an S-shape potential well just as shown Fig. 1. More electrons and ions can be trapped in the potential well as the LG laser screw-like drill into the plasma. The LG laser also works as a fan to blow the plasmas forward, which may enhance the absorption of the laser pulse (Fig. 5). Thus the total angular momentums of protons can be raised, just as shown in Fig. 7.

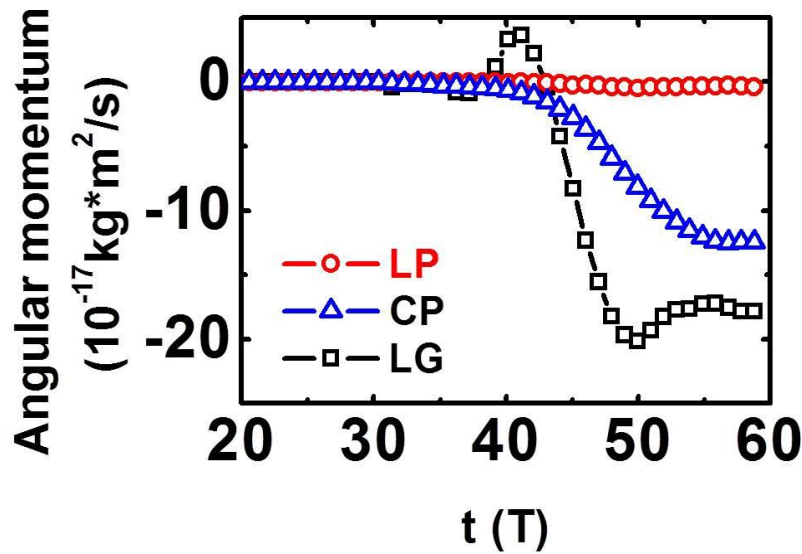

Figure 7    Total angular momentums of protons in the longitudinal direction for LG (squares, $a$ = 10), LP (circles, $a$ = 7.4), and CP (triangles, $a$ = 5.2) laser pulses at $t$ = 60$T$; $d$ = 1 μm, and $n_0$ = 4 $n_c$.

In conclusion, the particles rotation alignment in the tangential direction is realized with the use of an intense LG laser pulse. Compared with LP and CP lasers, the enhancement of the proton angular momentum in the longitudinal direction is obtained when an intense LP LG laser pulse irradiates on a thin foil. The PIC simulations show that the LG pulse can drill into the plasma like a screw. The angular momentum of the particles produced by the LG laser is enhanced compared with that produced by the LP and CP pulses with similar pulse energies. It is also found that electrons and protons are trapped and uniformly compressed in the dark central minimum of the doughnut LG pulse. LG laser has been generated by several techniques. For example, a high-order Hermite–Gaussian (HG) mode can be generated by inserting an intra-cavity cross-wire into a laser cavity. An LG laser can then be obtained using a mode converter on this HG laser[66]. A spiral phase plate[67] and a computer-generated hologram[49] may be used to generate the LG modes from a fundamental Gaussian mode (TEM$_{00}$). The intense LG laser exhibits potential applications in field of relativistic intensity, such as laser-driven plasma accelerators. Such LG laser can be applied in the TNSA or RPA experiments to generate a particle beam with angular momentum in the future.

**Methods**

The 3D simulations are performed with VORPAL, which is a relativistic, arbitrary, and dimensional hybrid plasma and beam simulation code. It includes the utilities for data analysis and scripts for data visualization. Particle-in-cell (PIC) algorithm is used in VORPAL to describe the kinetic plasma model. A charge-conserving current deposition algorithm is applied in electromagnetic limit to enable the integration of Maxwell's equations without any additional divergence correction. The instantaneous charge distribution is used to calculate Poisson's equation at every time step in the electrostatic limit.

In our simulation, a 40 fs $p$-polarized LG laser pulse is incident from the left side on the foil. The dimensionless peak amplitude of the incident laser pulse is $a_0 = eE_0/m_e \omega_L c = 5$ (the intensity is $I = 3.4 \times 10^{19}$ W/cm$^2$), where $\omega_L$ is the laser frequency, $\lambda = 1$ μm is the laser wavelength, $c$ is the light speed in a vacuum, and $m_e$ and $e$ are the electron mass and charge at rest. The radius of the laser is $W = 4$ μm and $W_x = 12$ μm. The laser front reaches the front surface of the foil at $t = 30T$ ($T = \lambda/c$). The foil thickness is $d = 1$ μm and the front surface of the foil is at $x = 30$ μm. The transverse range of the foil is $-14$ μm $< y <14$ μm and $-14$ μm $< z <14$ μm. The foil is assumed to be fully ionized into protons and electrons before the arrival of the main pulse. The foil density is $n_0 = 2 n_c$, where $n_c = \omega_L^2 m_e / 4\pi e^2$ is the critical density. A low-density step-like density profile is used to simplify the model and reduce the 3D PIC simulation time. The size of the simulation box is (60×60×60) μm, and the cell number is 600×600×600. Each cell is filled with 10 protons and 10 electrons.

Figure 1   (a) Schematic arrangement of the 3D PIC simulation. The distributions of electric fields in the ($y$, $z$) plane at (b) $x = 18$ μm, (c) $x = 18.2$ μm, (d) $x = 18.4$ μm, and (e) $x = 18.6$ μm.

Figure 2   Total angular momentums $m_e(yp_z - zp_y) + m_p(yp_z - zp_y)$ in the $x$ direction.

Figure 3   Angular momentums of particles at $60T$ for different laser amplitudes $a$. The initial foil density is $n_0 = 2\ n_c$ and the foil thickness is $d = 1$ μm.

Figure 4   LG laser interactions on the foils for (a–c) $a = 1$, (d–f) $a = 5$, and (g–i) $a = 10$. The distributions of the electric fields (first column), electron density (middle column), and ion density (third column) are shown at $t = 50T$.

Figure 5   Distributions of (a–c) electric fields, (d–f) electron densities, and (g–i) ion densities in the ($y$, $z$) plane at $x = 30$ μm (first column), $x = 30.4$ μm (second column), and $x = 30.8$ μm (third column) at $t = 50T$. The corresponding distributions are shown in Figs. 4(g) to 4(i).

Figure 6   Distributions of (a–d) electron densities and (e–h) ion densities in the ($y$, $z$) plane at $x = 30.5$ μm for different time $t = 35T$, $40T$, $45T$, and $50T$.

Figure 7   Total angular momentums of protons in the longitudinal direction for LG (squares, $a = 10$), LP (circles, $a = 7.4$), and CP (triangles, $a = 5.2$) laser pulses at $t = 60T$; $d = 1$ μm, and $n_0 = 2\ n_c$.

## Acknowledgements

This study was supported by the 973 Program (No. 2011CB808104), the National Natural Science Foundation of China (Nos. 11335013, 11125526, and 11305236), the International S&T Cooperation Program of China (No. 2011DFA11300), and the Shanghai Natural Science Foundation (No. 13ZR1463300).

## Author contributions

W.-P. W. and B.-F. S. contributed to all aspects of this work. B.-F. S., X.-M. Z. and Y. S. provided inspiring ideas to help W.-P. W. write the paper. L.-G. Z. helped in developing the plotting program in the 3D PIC simulations. Z- Z. X. gave some useful suggestions for this work. All authors discussed the results and commented on the manuscript.

## Additional information

Reprint and permission information is available online at www.nature.com/reprints. Correspondence and requests for materials should be addressed to W.-P. W.

## Competing financial interests

The authors declare no competing financial interests.